# Solar Neutrinos: Where We Are, Where We Are Going[1]

John N. Bahcall

Institute for Advanced Study, School of Natural Sciences,

Princeton, NJ 08540

## Abstract

This talk answers a series of questions. Why study solar neutrinos? What does the combined standard model (solar plus electroweak) predict for solar neutrinos? Why are the calculations of neutrino fluxes robust? What are the three solar neutrino problems? What have we learned in the first thirty years of solar neutrino research? For the next decade, what are the most important solvable problems in the physics of solar neutrinos? What are the most important problems in the astrophysics of solar neutrinos?

## 1. Why Study Solar Neutrinos?

Astronomers study solar neutrinos for different reasons than physicists. For astronomers, solar neutrino observations offer an opportunity to test directly the theories of stellar evolution and of nuclear energy generation. With neutrinos one can look into the interior of a main sequence star and observe the nuclear fusion reactions that are ultimately responsible for starlight via the general reaction:

---

$$4{}^1\mathrm{H} \longrightarrow {}^4\mathrm{He} + 2e^+ + 2\nu_e. \tag{1}$$

The optical depth of the sun for a typical neutrino produced by nuclear fusion is $\sim 10^{-9}$, about 20 orders of magnitude smaller than the optical depth for a typical optical photon. As we shall see, solar neutrino experiments constitute quantitative, well-defined tests of the theory of stellar evolution. Since stellar evolution theory is used widely in interpreting astronomical observations, direct tests of this theory are of importance to astronomers.

Table 1 shows the principal nuclear reactions that accomplish Eq. (1) via the proton-proton chain of reactions. In what follows, we shall refer often to the reactions listed in this table.

Table 1: The Principal Reactions of the $pp$ Chain

| Reaction | Reaction Number | Neutrino Energy (MeV) |
|---|---|---|
| $p + p \to {}^2\mathrm{H} + e^+ + \nu_e$ | 1 | 0.0 to 0.4 |
| $p + e^- + p \to {}^2\mathrm{H} + \nu_e$ | 2 | 1.4 |
| ${}^2\mathrm{H} + p \to {}^3\mathrm{He} + \gamma$ | 3 | |
| ${}^3\mathrm{He} + {}^3\mathrm{He} \to {}^4\mathrm{He} + 2p$ | 4 | |
| or | | |
| ${}^3\mathrm{He} + {}^4\mathrm{He} \to {}^7\mathrm{Be} + \gamma$ | 5 | |
| then | | |
| $e^- + {}^7\mathrm{Be} \to {}^7\mathrm{Li} + \nu_e$ | 6 | 0.86, 0.38 |
| ${}^7\mathrm{Li} + p \to {}^4\mathrm{He} + {}^4\mathrm{He}$ | 7 | |
| or | | |
| $p + {}^7\mathrm{Be} \to {}^8\mathrm{B} + \gamma$ | 8 | |
| ${}^8\mathrm{B} \to {}^8\mathrm{Be} + e^+ + \nu_e$ | 9 | 0 to 14 |

Solar neutrinos are of interest to physicists because they can be used to perform unique particle physics experiments. For some of the theoretically most interesting ranges of masses and mixing angles, solar neutrino experiments are more sensitive tests for neutrino transformations in flight than experiments that can be carried out with laboratory sources. The reasons for this exquisite sensitivity are: 1) the great distance between the accelerator (the solar interior) and the detector (on earth); 2) the relatively low energy (MeV) of solar neutrinos; and 3) the enormous path length of matter ($\sim 10^{11}$gm cm$^{-2}$ ) that neutrinos must pass through on their way out of the sun.

One can quantify the sensitivity of solar neutrinos relative to laboratory experiments by considering the proper time that would elapse for a finite-mass neutrino in flight between the point of production and the point of detection. The elapsed proper time is a measure of the opportunity that a neutrino has to transform its state and is proportional to the ratio, $R$, of path length divided by energy:

$$\text{Proper Time} \propto R = \frac{\text{Path Length}}{\text{Energy}}. \tag{2}$$

Future accelerator experiments with multi-GeV neutrinos may reach a sensitivity of $R = 10^2$ Km/(1 GeV). Reactor experiments have already almost reached a level of sensitivity of $R = 10^2$ Km/(1 GeV) for neutrinos with MeV energies. Solar neutrino experiments, because of the enormous distance between the source (the center of the sun) and the detector (on earth) and the relatively low energies (1 MeV to 10 MeV) of solar neutrinos involve much larger values of neutrino proper time,

$$R(\text{solar}) = \frac{10^8}{10^{-3}} \left( \frac{\text{Km}}{\text{GeV}} \right) \sim 10^{11} \left( \frac{\text{Km}}{\text{GeV}} \right). \tag{3}$$

Because of the long proper time that is available to a neutrino to transform its state, solar neutrino experiments are sensitive to very small neutrino masses and transition matrix elements which can cause neutrino oscillations in vacuum. Quantitatively,

$$m_\nu(\text{solar level of sensitivity}) \sim 10^{-6}\text{eV to } 10^{-5}\text{eV} \quad (\text{vacuum oscillations}), \tag{4}$$

provided the electron-neutrino that is created by beta-decay contains appreciable portions of at least two different neutrino mass eigenstates (i. e., the neutrino mixing angle is relatively large). Laboratory experiments have achieved a sensitivity to electron neutrino masses of order one eV. Over the next several years, the sensitivity of the laboratory experiments should be improved by an order of magnitude or more.

Resonant neutrino oscillations, which may be induced by neutrino interactions with electrons in the sun (the famous MSW effect), can occur even if the electron neutrino is almost entirely composed of one neutrino mass eigenstate (i. e., even if the mixing angles between $e - \mu$ and $e - \tau$ neutrinos are tiny). Standard solar models indicate that the sun has a high central density, $\rho(\text{central}) \sim 1.5 \times 10^2$ gm cm$^{-3}$, which (at least in principle) allows even very low energy ($< 1$ MeV) electron-type neutrinos to be resonantly converted (to the more difficult to detect $\mu$ or $\tau$ neutrinos) by the MSW effect. Also, the column density of matter that neutrinos must pass through is large: $\int \rho dr \approx 2 \times 10^{11}$ gm cm$^{-2}$. The corresponding parameters for terrestrial, long-baseline experiments are: a typical density of 3 gm cm$^{-3}$, and an obtainable column density of about $2 \times 10^8$ gm cm$^{-2}$.

Given the above solar parameters, the planned and operating solar neutrino experiments are sensitive to neutrino masses in the range

$$10^{-4} eV \;\lesssim\; m_\nu \;\lesssim\; 10^{-2} eV, \tag{5}$$

via matter-induced resonant oscillations (MSW effect).

The range of neutrino masses given by Eq. (4) and Eq. (5) is included in the range of neutrino masses that are suggested by attractive particle-physics generalizations of the standard electroweak model, including left-right symmetry, grand-unification, and supersymmetry.

Both vacuum neutrino oscillations and matter-enhanced neutrino oscillations can change electron-type neutrinos to the more difficult to detect muon or tau neutrinos. In addition, the likelihood that a neutrino will have its type changed may depend upon its energy, affecting the shape of the energy spectrum of the surviving electron-type neutrinos. Future solar neutrino experiments will measure the ratio of the number of electron-type neutrinos to the total number of solar neutrinos (via neutral current reactions) and will also measure the shape of the electron-type energy spectrum (via charged current absorption and by neutrino-electron scattering). These measurements will test the simplest version of the standard electroweak model (in which neutrinos are massless and do not oscillate); these tests are independent of solar model physics.

## 2.  What Does the Combined Standard Model Tell Us About Solar Neutrinos?

In this section, I will describe the combined standard model (standard solar model and standard electroweak theory) that is used to decide if solar neutrino experiments have revealed something unexpected. Then I will present the calculated solar neutrino spectrum as predicted by the standard model.

### 2.1.  The Combined Standard Model

In order to interpret solar neutrino experiments, one must have a quantitative theoretical model. Unlike many other aspects of astronomy in which one can make important discoveries by identifying new classes of objects (such as quasars, or $x-$ray sources or $\gamma-$ ray sources), solar neutrino research requires a reliable theoretical model for comparison with the observations in order to determine whether one has found something surprising. Our physical intuition is not yet sufficiently advanced to know if we should be surprised by $10^{-2}$, by $10^{0.0}$, or by $10^{+2}$ neutrino-induced events per day in a chlorine tank the size of an Olympic swimming pool.

I will use the most conservative model for comparison with experiments, the combined standard model, unless explicitly stated otherwise. The combined standard model is the standard model of solar structure and evolution and the standard electroweak model of particle physics.

A solar model is required in order to predict the number of neutrinos created in a given energy range per unit of time. On a fundamental level, a solar model is required in order to predict the rate of nuclear fusion by the $pp$ chain (shown in Table 1 and discussed below) and the rate of fusion by the CNO reactions (originally favored by H. Bethe in his epochal study of nuclear fusion reactions).

In our discussion, I will assume a result common to all modern solar models, namely, that the CNO reactions contribute only a very small fraction of the luminosity of the sun. Although the dominance of the $pp$ chain is often taken for granted in theoretical analyses of solar neutrino experiments, it is not an *a priori* obvious result. We shall come back to this fundamental prediction of solar models in the discussion given in §5.1, where we review what has been learned about astronomy from solar neutrino experiments.

A precise solar model is required to calculate accurately which nuclear reaction occurs more often at the two principal branching points of the $pp$ fusion chain. Referring to Table 1, the branching points occur between reactions 4 and 5 and between reactions 6 and 8. If the $pp$ chain is terminated by reaction 4, only low-energy ($< 0.4$ MeV) $pp$ neutrinos are produced, whereas if the termination occurs via reaction 5 higher-energy $^7$Be and $^8$B neutrinos are created. The ratio of the rates for reaction 6 and reaction 8 determines how often $^7$Be neutrinos (two lines: 0.86 MeV and 0.38 MeV) are produced rather than the rare, but more easily detected $^8$B neutrinos (maximum energy $\sim 14$ MeV) are produced.

The competition between reactions 4 and 5, and between reactions 6 and 8, determines the energies of the emitted solar neutrinos. The predicted rates in solar neutrino experiments depend sensitively upon the relative frequencies of these crucial reactions. Fortunately, the theoretical uncertainties in the predicted neutrino fluxes are not very large; they vary between about 1% to about 17%, depending upon the neutrino source in question. The rates for individual detectors are determined by the energy spectrum, by the neutrino type of the incoming solar neutrinos, and by the energy-dependence of the interaction cross sections of the different detectors. As we shall see in the later discussion (cf. §4), one can largely avoid the dependence of the predictions upon solar models by comparing the results of experiments that have different energy sensitivities. These intercomparisons between experiments primarily test standard electroweak theory.

A particle physics model is required to predict what happens to the neutrinos after they are created. I will use the simplest version of the standard electroweak model according to which nothing happens to the neutrinos after they are created in the interior of the sun. In this theory, neutrinos are massless and neutrino flavor (electron type) is conserved. The standard electroweak model has had many successes in precision laboratory tests; modifications of this theory will be accepted only if incontrovertible experimental evidence forces a change.

## 2.2. The Solar Neutrino Spectrum

Figure 1 shows the calculated neutrino spectrum for the most important neutrino sources from the sun. I will discuss briefly the $pp$ (and $pep$) neutrinos, the $^7$Be neutrinos, and the $^8$B neutrinos. I will concentrate on the reliability of the predictions and will indicate the role of each of these neutrinos in the ongoing experiments.

The dominant source of solar neutrinos is produced by the first reaction listed in Table 1, the basic $pp$ reaction ($p + p \longrightarrow D + e^+ + \nu_e$), which creates neutrinos with energies less than 0.4 MeV. Most of the nuclear energy that emerges as sunlight begins with this reaction. The theoretical uncertainty in the $pp$ neutrino flux is less than 1%. Of the planned and operating solar neutrino experiments, only the GALLEX and SAGE gallium experiments have energy thresholds low enough to detect the $pp$ neutrinos. (About 0.2% of the $pp$ fusions are believed to occur via the $pep$ reaction, the second reaction in Table 1. This reaction produces a neutrino line that contributes a small part of the calculated event rate in the chlorine and gallium experiments.)

The next most important source of neutrinos is from the $^7$Be neutrino line at 0.86 MeV, which is produced by reaction 6 of Table 1. About 15% of the solar luminosity is produced by reactions which go through this channel; the uncertainty in the neutrino flux is about 7% . The $^7$Be neutrinos contribute significantly, according to standard model calculations, to the chlorine and the gallium experiments, but are too low in energy to be detected in the Kamiokande experiment. In an experiment under development called BOREXINO, $^7$Be neutrinos will be detected by the unique signature they produce in scintillation light caused by neutrino-electron scattering.

The $^8$B neutrino flux, produced by reaction 9 of Table 1, is tiny, about $10^{-4}$ of the flux of $pp$ neutrinos. However, the $^8$B neutrinos are crucial for solar neutrino physics and astronomy. Because of their high energy ($\sim$ 10 MeV, which takes advantage of a superallowed transition from the ground-state of chlorine to an excited state of argon), $^8$B neutrinos dominate the predicted capture rate for the chlorine experiment. They are also the only significant source of neutrinos above the energy threshold in the water Cherenkov experiment, Kamiokande. The SNO and Superkamiokande experiments, both of which are under development and which should begin producing results in 1996, also observe only $^8$B neutrinos. Unfortunately, the theoretical uncertainty in the predicted $^8$B neutrino flux is relatively large, about 17%.

### 3. Why Are the Predicted Neutrino Fluxes Robust?

The predicted event rates in the different solar neutrino experiments have been remarkable stable over the past 25 years. The published estimate in 1968, which accompanied the first report by Davis and his collaborators of measurements with the chlorine experiment, was $7.5 \pm 1.0$ SNU; the most recent and detailed calculation gives in 1995 a predicted rate of $9.3 \pm 1.3$ SNU . The theoretical errors are intended to be as close as possible to effective $1\sigma$ errors; they are obtained by carrying out detailed calculations using $1\sigma$ uncertainties on all the measured input data and, for the less important theoretical errors, by taking the extreme range of theoretical calculations to be $3\sigma$ uncertainties.

There are three reasons that the theoretical calculations of the neutrino fluxes are robust: 1) the availability of precision measurements and precision calculations of input data; 2) the connection between neutrino fluxes and the measured solar luminosity; and 3) the measurement of the helioseismological frequencies of the solar pressure-mode ($p$-mode) eigenfrequencies.

Over the past three decades, many hundreds of researchers have performed precision measurements of crucial input data including nuclear reaction cross sections and the abundances of the chemical elements on the solar surface. Many other researchers have calculated accurate opacities, equations of state, and weak interaction cross sections. By now, these input data are relatively precise and their uncertainties are quantifiable.

The solar neutrino fluxes and the solar luminosity both depend upon the rates of the nuclear fusion reactions in the solar interior. Since we know experimentally the solar luminosity (to an accuracy of about 0.4%), the calculated neutrino fluxes are strongly constrained by the fact that the standard solar models must yield precisely the measured solar luminosity.

Thousands of $p$-mode helioseismological frequencies have been measured to an accuracy of 1 part in $10^4$. The best standard solar models are required to reproduce all of these $p$-mode frequencies to an accuracy of better than one part in a thousand. In fact, standard solar models are in agreement with the measured helioseismological frequencies to a high level of precision without any special adjustments of the parameters.

The calculated solar neutrino fluxes are, after thirty years of intense study, known to relatively high accuracy because of the many precise measurements and calculations of input data, because of the strong constraint imposed on the models by the measured total solar luminosity, and because of the important tests of solar structure that are provided by helioseismological measurements.

## 4.   What Are the Three Solar Neutrino Problems?

I will compare in this section the predictions of the combined standard model with the results of the operating solar neutrino experiments. We will see that this comparison leads to three different discrepancies between the calculations and the observations, which I will refer to as the three solar neutrino problems.

Figure 2 shows the measured and the calculated event rates in the four ongoing solar neutrino experiments. This figure reveals three discrepancies between the experimental results and the expectations that are based upon the combined standard model. As we shall see, only the first of these discrepancies depends sensitively upon predictions of the standard solar model.

## 4.1. Calculated Versus Observed Chlorine Rate

The first solar neutrino experiment to be performed was the chlorine radiochemical experiment, which detects electron-type neutrinos more energetic than 0.81 MeV. After more than twenty five years of the operation of this experiment, the measured event rate is $2.55 \pm 0.25$ SNU, which is a factor of about 3.6 less than is predicted by the most detailed theoretical calculations, $9.3^{+1.2}_{-1.4}$ SNU. A SNU is a convenient unit to describe the measured rates of solar neutrino experiments: $10^{-36}$ interactions per target atom per sec. Most of the predicted rate in the chlorine experiment is from the rare, high-energy $^8$B neutrinos, although the $^7$Be neutrinos are also expected to contribute significantly. According to standard model calculations, the *pep* neutrinos and the CNO neutrinos (for simplicity, not discussed here) are expected to contribute less than a SNU to the total event rate.

This discrepancy between the calculations and the observations for the chlorine experiment was, for more than two decades, the only solar neutrino problem. I shall refer to the chlorine disagreement as the "first" solar neutrino problem.

## 4.2. Incompatibility of Chlorine and Water (Kamiokande) Experiments

The second solar neutrino problem results from a comparison of the measured event rates in the chlorine experiment and in the Japanese pure-water experiment, Kamiokande. The water experiment detects higher-energy neutrinos, those with energies above 7.5 MeV, by neutrino-electron scattering: $\nu \; + \; e \; \longrightarrow \; \nu' \; + \; e'$. According to the standard solar model (see also Table 1), $^8$B beta-decay is the only important source of these higher-energy neutrinos.

The Kamiokande experiment shows that the observed neutrinos come from the sun. The electrons that are scattered by the incoming neutrinos recoil predominantly in the direction of the sun-earth vector; the relativistic electrons are observed by the Cherenkov radiation they produce in the water detector.

In addition, the Kamiokande experiment measures the energies of individual scattered electrons and therefore provides information about the energy spectrum of the incident solar neutrinos. The observed spectrum of electron recoil energies is consistent with that expected from $^8$B neutrinos. However, small angle scattering of the recoil electrons in the water prevents the angular distribution from being determined well on an event-by-event basis, which limits the constraints the experiment places on the incoming neutrino spectrum.

The event rate in the Kamiokande experiment is determined by the same high-energy $^8$B neutrinos that are expected, on the basis of the combined standard model, to dominate the event rate in the chlorine experiment. I have shown that solar physics changes the shape of the $^8$B neutrino spectrum by less than 1 part in $10^5$. Therefore, we can calculate the rate in the chlorine experiment that is produced by the $^8$B neutrinos observed in the Kamiokande experiment (7.5 MeV threshold energy). This partial ($^8$B) rate in the chlorine experiment is $3.2 \pm 0.45$ SNU, which exceeds the total observed chlorine rate of $2.55 \pm 0.25$ SNU. (This problem is analogous to the familiar astronomical problem with the Hubble constant, in which the age of the oldest stars appears to exceed the age of the universe, at least when the most popular current values are used.)

Comparing the rates of the Kamiokande and the chlorine experiments, one finds that the net contribution to the chlorine experiment from the $pep$, $^7$Be, and CNO neutrino sources is negative: $-0.66 \pm 0.52$ SNU. (The calculated rate from $pep$, $^7$Be, and CNO neutrinos is 1.9 SNU.) The apparent incompatibility of the chlorine and the Kamiokande experiments is the "second" solar neutrino problem.

### 4.3.  Gallium Experiments: No Room for $^7$Be Neutrinos

The results of the gallium experiments, GALLEX and SAGE, constitute the third solar neutrino problem. The average observed rate in these two experiments is 74 SNU, which is essentially fully accounted for in the standard model by the theoretical rate of 73 SNU that is calculated to come from the basic $pp$ and $pep$ neutrinos (with only a 1% uncertainty in the standard solar model $pp$ flux). The $^8$B neutrinos, which are observed above 7.5 MeV in the Kamiokande experiment, must also contribute to the gallium event rate. Using the standard model shape for the spectrum of $^8$B neutrinos and normalizing to the rate observed in Kamiokande, $^8$B contributes another 7 SNU, unless something happens to the lower-energy neutrinos after they are created in the sun. (The predicted contribution is 16 SNU on the basis of the standard model.) Given the measured rates in the gallium experiments, there is no room for the additional $34 \pm 4$ SNU that is expected from $^7$Be neutrinos on the basis of 13 standard solar models calculated by different groups using different input data and different stellar evolution codes.

The seeming exclusion of everything but $pp$ neutrinos in the gallium experiments is the "third" solar neutrino problem. This problem is essentially independent of the previously-discussed solar neutrino problems, since it depends upon the $pp$ neutrinos that are not observed in the other experiments and whose calculated flux is approximately model-independent (if the general scheme of the $pp$ chain shown in Table 1 is correct).

The missing $^7$Be neutrinos cannot be explained away by any change in solar physics. The $^8$B neutrinos that are observed in the Kamiokande experiment are produced in competition with the missing $^7$Be neutrinos; the competition is between reaction 6 and reaction 8 in Table 1. Solar model explanations that reduce the predicted $^7$Be flux reduce much more (too much) the predictions for the observed $^8$B flux.

The flux of $^7$Be neutrinos, $\phi(^7\text{Be})$, is independent of measurement uncertainties in the cross section for the nuclear reaction $^7\text{Be}(p, \gamma)^8\text{B}$ ; the cross section for this proton-capture reaction is the most uncertain quantity that enters in an important way in the solar model calculations. The flux of $^7$Be neutrinos depends upon the proton-capture reaction only through the ratio

$$\phi(^7\text{Be}) \;\propto\; \frac{\text{R}(e)}{\text{R}(e) + \text{R}(p)}, \qquad (6)$$

where $R(e)$ is the rate of electron capture by $^7$Be nuclei and $R(p)$ is the rate of proton capture by $^7$Be. With standard parameters, solar models give $R(p) \;\approx\; 0.001 R(e)$. Therefore, one would have to increase the value of the $^7\text{Be}(p, \gamma)^8\text{B}$ cross section by more than two orders of magnitude over the current best-estimate (which has an estimated uncertainty of about 10%) in order to affect significantly the calculated $^7$B solar neutrino flux. The required change in the nuclear physics cross section would also increase the predicted neutrino event rate by more than a hundred in the Kamiokande experiment, making that prediction completely inconsistent with what is observed. (From time to time, papers have been published claiming to solve the solar neutrino problem by artifically changing the rate of the $^7$Be electron capture reaction. Equation Eq. (6) shows that the flux of $^7$Be neutrinos is actually independent of the rate of the electron capture reaction to an accuracy of better than 1 %.)

I conclude that either: 1) at least three of the four operating solar neutrino experiments (the two gallium experiments plus either Chlorine or Kamiokande) give misleading results, or 2) physics beyond the standard electroweak model is required to change the neutrino spectrum or flavor content after the neutrinos are produced in the center of the sun.

## 5.   What Have We Learned?

No solar-model solution has been found that explains the results of the four existing solar neutrino experiments. Many particle-physics solutions have been proposed that can explain the existing data.

In this section, I will summarize the main lessons that have been learned from the first thirty years of solar neutrino research. I will first review the progress in astrophysics and then outline briefly the developments in physics.

### 5.1.   Learned About Astronomy

The chlorine solar neutrino experiment was proposed in 1964 as a practical test of solar model calculations in a pair of back-to-back (Physical Review Letters, Bahcall and Davis) papers. The only motivation presented in those two papers for performing the chlorine experiment was "...to see into the interior of a star and thus verify directly the hypothesis of nuclear energy generation in stars."

What have we learned by direct experiments about nuclear energy generation in stars? How does our 1964 understanding compare with the results of the solar neutrino experiments? Table 2 summarizes the six principal predictions that were made (or which were implicit in the theory) in 1964 and compares those predictions with the results of the four ongoing solar neutrino experiments.

The neutrinos were predicted to originate in the solar interior; the direction of origin of the neutrinos has been verified by detecting neutrino-electron scattering (as was also suggested in a futuristic paper in 1964) in the Kamiokande experiment. The rates of the four operating experiments are in semi-quantitative agreement with the predictions; the ratios of the observed to the predicted rates are 0.3 (chlorine), 0.5 (water-Kamiokande), and 0.5 (gallium, average). This agreement is better than any of us dared hope for in 1964, especially since the dominant neutrino flux (from $^8$B beta-decay) for the first two experiments depends upon the central temperature of the sun as approximately the 20th power of the central temperature. The energy range of the dominant neutrinos was predicted to be from 0 MeV to 14 MeV, which is consistent with the observations from the Kamiokande experiment. These observations do not give detailed information on the incoming solar neutrino energy spectrum, but they do show that neutrinos exist in the expected energy range (at least above 7.5 MeV) and that no events have been observed at energies beyond the maximum energy resulting from $^8$B beta-decay. Standard models predict that the neutrino fluxes are constant in time except for a small seasonal variation. (The Kelvin-Helmholtz cooling time for the solar interior is about $10^7$ years.) The consensus, but not unanimous view of the experimentalists, based upon the available data from the pioneering four solar neutrino experiments (all of which have rather low counting rates) is that the neutrino fluxes are indeed constant in time. Standard solar models predict that the sun shines almost entirely via the *pp* chain of nuclear fusion reactions, rather than the CNO reactions originally emphasized by Bethe. If the CNO reactions were dominant, the event rates in solar neutrino experiments could be calculated precisely and these rates, as shown in Table 2, differ from the observed rates by more than an order of magnitude. Finally, if we crudely characterize the rate of the $^8$B neutrino emission by its approximate dependence upon the central temperature of the solar model, then the central temperature of the solar model agrees with the value obtained from the experimental rates to an accuracy of about 4% or better.

Table 2: Nuclear energy generation. Predictions vs. observations.

| Predicted | Observed |
| --- | --- |
| Direction: the Sun | o.k. |
| Rates (4 expts.) | $\sim$ Predicted Rates (within factor of a few) |
| 0–14 MeV | < 14 MeV |
| Constant in time (except seasonal) | $\sim$ o.k. |
| If CNO : $\begin{cases} \text{Cl}: & 28 \text{ SNU} \\ \text{Ga}: & 610 \text{ SNU} \\ \text{H}_2\text{O}: & 0.0 \end{cases}$ | 2.6 SNU<br>74 SNU<br>0.44 Standard Model |
| $T_{\text{model}}(r = 0) = 16 \times 10^6$ K | $T(^8\text{B})/T_{\text{model}} \gtrsim 0.96$ |

The four ongoing solar neutrino experiments have shown directly that the sun shines by nuclear fusion reactions, achieving the original goal. Quantitative improvements in the tests shown in Table 2 will occur with the next generation of experiments. But, the most important qualitative result has been established: neutrinos have been observed from the interior of the sun in approximately the number and with the energies expected.

In the three-decade long struggle to improve solar models in order to calculate more accurate solar neutrino fluxes, we have obtained a greater understanding of solar structure. The theoretical models have gradually been refined as improved input data, more accurate physical descriptions, and more precise numerical techniques have been employed. Perhaps most importantly, the complimentary field of helioseismology has been developed and now provides precise data that determine the sound velocity over most of the solar interior; these beautiful measurements are used to test and to refine the standard solar model.

Further improvements in the solar model are desirable and important, but the quantitative agreement, typically better than a part in $10^3$, between the calculated eigenfrequencies of pressure modes and the measured (helioseismological) frequencies provides strong evidence for the basic correctness of the standard solar model.

## 5.2. Learned About Physics

To perform more precise tests of the astrophysical predictions, we will have to learn what happens to neutrinos after they are produced in the interior of the sun. Theoretical physicists have fertile imaginations; they have provided us with a rich smorgasbord of explanations based upon new particle physics, including: vauum neutrino oscillations, resonant oscillations in matter (the MSW effect), resonant magnetic-moment transitions, sterile neutrinos, neutrino decay, and violation by neutrinos of the equivalence principle. Most of these explanations can account for the existing experimental data if either two or three neutrinos are involved in the new physics beyond the standard electroweak model. All of these explanations, and others that I have not listed, can account for the existing data on solar neutrino experiments without conflicting with established laws of physics or with other experimental constraints.

The number of proposed particle physics explanations exceeds the diagnostic power of the existing solar neutrino experiments. I think it is unlikely that the next generation of solar neutrinos experiments will be able to eleminate all but one possible particle physics explanation. But, I hope that the powerful new experiments (SNO, Superkamiokande, and BOREXINO) will, together with the four operating experiments, point us in the direction of one of the previously-proposed explanations.

So, what have we learned about particle physics? We have learned that a number of particle physics explanations are consistent with the data obtained from the first four solar neutrino experiments. Perhaps most importantly, we have learned that an elegant, conservative extension of the standard electroweak theory, the MSW effect, can describe all of the existing experimental information on solar neutrinos if the electron neutrino is mixed with another neutrino that has a finite mass,

$$m_\nu \approx 0.003 \text{ eV}. \tag{7}$$

The MSW theory is not proven, but it is a beautiful idea. I think it would be a disgrace if Nature failed to make use of this marvelous possibility.

## 6.   What Next?

In this section, I will summarize the problems, first in physics and then in astronomy, that are likely to be solved in the next decade or two.

## 6.1.   Solvable Problems in Physics

The fundamental goal of physics research on solar neutrinos is to measure the energy spectrum and flavor content as a function of time of the solar neutrino flux. We want to know how many neutrinos reach the earth with a given energy and with a given flavor (i.e., neutrino type: electron, muon, or tau), all as a function of time. Because of some exotic particle physics possibilities, we also want to know if the solar neutrino flux contains any anti-neutrinos.

The standard model predicts that the energy spectrum of neutrinos from any given neutrino source, e. g., from $^8$B beta-decay, will be the same to high accuracy as the energy spectrum inferred from terrestrial laboratory measurements. In the standard electroweak theory, only massless electron-type neutrinos are created in nuclear beta decay or nuclear fusion reactions. Standard electroweak theory predicts that the solar neutrinos produced by nuclear fusion reactions are all $\nu_e$, not $\nu_\mu$ or $\nu_\tau$. (According to MSW and vacuum oscillation theories, neutrinos created in nuclear beta-decay or nuclear fusion reactions are linear combinations of different neutrino types and at least one neutrino mass is non-zero.) Finally, the total amount of thermal energy in the solar interior implies that the neutrino fluxes will be constant in time (for time scales less than $10^7$ years) except for the seasonal dependences caused by the earth's orbital eccentricity. Any departure from these expectations will be a signal for physics beyond the standard electroweak model.

I will now list six specific problems in the physics of solar neutrino research that we can expect to be solved in the next decade or so.

• **The ratio of $\nu_e$ to $\nu_{\text{total}}$** . The electron-type neutrinos, $\nu_e$, that are created in the interior of the sun will all remain $\nu_e$ if the simplest version of the standard electroweak theory is correct. If neutrino oscillations occur, vacuum or matter-induced (MSW effect) oscillations, then the total number of solar neutrinos observed at earth will exceed the number that are present as electron-type neutrinos only.

The ratio of $\nu_e$ to $\nu_{\text{total}}$ can be determined by measuring the ratio of the cross sections for two different reactions, one that occurs only with electron-type neutrinos and one that will occur with equal probability independent of the type of neutrino. This ratio will be measured for the first time in the SNO solar neutrino experiment[2], which utilizes a kiloton of heavy water, via the two reactions: $\nu_e + {}^2\text{H} \longrightarrow p + p + e$ (only $\nu_e$); and

---

[2] If non-interacting (sterile) neutrinos, $\nu_{\text{sterile}}$, exist in nature, they will not be detected in this experiment. In principle, the solution of the solar neutrino problems could involve $\nu_e \to \nu_{\text{sterile}}$.

$\nu_{\text{total}} + {}^2\text{H} \longrightarrow p + n + \nu_{\text{total}}.$

• **Shape of the ${}^8$B neutrino energy spectrum.** The shape of this spectrum is independent of any aspect of solar model physics to an accuracy of 1 part in $10^5$. The shape is determined empirically from laboratory nuclear physics measurements.

The shape of the ${}^8$B neutrino energy spectrum can be determined by measuring the energy spectrum of electrons created in the reaction $\nu_e + {}^2\text{H} \longrightarrow p + p + e$, which will be done in the SNO experiment. Important information about the energy spectrum will also be obtained by the Superkamiokande electron-scattering experiment. Superkamiokande will measure accurately the energy spectrum of recoiling electrons produced by neutrino-electron scattering in pure water.

• **Flux of ${}^7$Be neutrinos.** The simplest particle-physics interpretation of the four operating solar neutrino experiments implies that the flux of ${}^7$Be neutrinos is greatly reduced with respect to the value predicted by the standard solar model. I do not know of any proposed modification of solar physics that could explain a greatly reduced ${}^7$Be neutrino flux and at the same time be consistent with only a factor of two reduction in the ${}^8$B neutrino flux (as observed by the Kamiokande experiment). Therefore, a much reduced ${}^7$Be neutrino flux would be a strong signal for new weak interaction physics.

A measurement of the ${}^7$Be neutrino flux is also required to interpret the measured event rates in the existing chlorine and gallium experiments. Since the chlorine and gallium experiments are radiochemical, they do not give any indication of the energy (above threshold) of the neutrino that initiates the reaction.

A first measurement of the flux of ${}^7$Be neutrinos will be carried out by the BOREXINO collaboration in the Gran Sasso underground laboratory using electron-neutrino scattering in an organic scintillator.

•**Time Dependence of the Neutrino Fluxes.** All of the proposed experiments have expected event rates that are one to two orders of magnitude larger than the operating solar neutrino experiments. The tests for time-dependence with the existing data have yielded only marginally suggestive results. Typical event rates with existing experiments are in the range of 25 to 50 events per year, whereas the expected event rates in the planned experiments are typically of order a few thousand events per year.

With these future experiments, it will be possible to test with high precision the standard model prediction that the solar neutrino event fluxes are independent of time. In addition, it will be possible to measure the 7% peak-to-peak seasonal dependence caused by the eccentricity of the earth's orbit. One can also search with high accuracy for the strong seasonal dependences predicted by explanations involving vacuum neutrino oscillations, for which the oscillation wavelength is tuned to give a large effect by being set approximately equal to an integral multiple of the astronomical unit. MSW theory also predicts, for certain choices of the parameters, that there will be a strong day-night effect. This effect occurs because muon or tau neutrinos are reconverted to electron neutrinos as they pass through the earth at night on their way to neutrino detectors on the far side from the sun.

• **Proton-proton Neutrino Flux and Energy Spectrum.** The dominant neutrino flux that is created in the sun is the low energy flux of neutrinos from the basic $pp$ reaction. Because of their low energy ($< 0.4\text{MeV}$), these neutrinos are unobservable in the chlorine, Kamiokande, Superkamiokande, SNO, BOREXINO, and Indium experiments. Among the planned or operating experiments, only the gallium experiments, GALLEX and SAGE, have a sufficiently low threshold energy to detect $pp$ neutrinos. The gallium experiments detect neutrinos radiochemically; they do not measure the energy of the neutrinos that cause the conversion of $^{71}$Ga to $^{71}$Ge. Therefore, we have no experimental way at present of determining how much of the observed event rate in the gallium experiments is due to $pp$ neutrinos and how much is due to $^7$Be, CNO, or $^8$B neutrinos.

Two experiments have been proposed recently that are potentially capable of detecting and measuring the energies of individual events caused by $pp$ neutrinos. These experiments both would use cold helium. The HERON detector uses ballistic phonon propagation in liquid helium maintained in the superfluid state. The HELLAZ detector uses a high-pressure helium gas in a time-projection chamber.

The development of a practical detector that can measure the rate and the energy spectrum of the $pp$ neutrinos is an exciting technical challenge and of fundamental importance to both physics and astronomy. All of the proposed physics solutions make definitive predictions about what happens to the low-energy $pp$ neutrinos.

• **Refine Nuclear-physics Parameters.** Over the past three decades, many precise, difficult, and beautiful nuclear physics experiments have been performed in order to determine the rates of solar fusion reactions with the accuracy required for solar-model calculations of neutrino fluxes. It is important to check these measurements with the improved experimental techniques that are now available.

The most important nuclear physics experiment to perform in connection with solar neutrino research is a measurement of the $^7$Be $+$ $p$ $\longrightarrow$ $^8$B$^*$ $+$ $\gamma$ cross section (reaction 8 of Table 1). The predicted rates in the Kamiokande, Superkamiokande, and SNO solar neutrino experiments are proportional to the low-energy rate of this reaction and the dominant contribution in the standard model prediction of the rate of the chlorine experiment is also proportional to the rate of this reaction. There are a number of very beautiful experiments in which the rate of the $p(^7$Be$, \gamma)^8$B reaction has been measured using a radioactive target of $^7$Be and a beam of protons. It would be most informative to reverse the usual experimental situation and to use a gaseous target of protons and a beam of $^7$Be; this reversal would involve different systematic uncertainties, which are often most important source of errors in difficult experiments. Measuring the cross section for the $^7$Be$(p, \gamma)^8$B reaction with a $^7$Be beam is, in my view, the most important experiment to be performed in nuclear astrophysics.

The fundamental goal of physics experiments with solar neutrinos is to measure, or set stringent limits on, the elementary properties of neutrinos, especially their masses and mixing angles. It seems likely that we will make important progress towards this goal in the next decade.

## 6.2. Solvable Problems in Astronomy

The fundamental goal of solar neutrino astronomy is to determine the rates of different nuclear fusion reactions in the solar interior. Neutrino fluxes created by the different nuclear sources are the signatures of the fusion reactions. We must know what happens to the neutrinos after they are created in order to infer the created neutrino energy spectrum from the measured neutrino energy spectrum.

With one exception, progress in solar neutrino astronomy is held hostage to progress in particle physics. As discussed in the previous subsection, it seems likely that we will learn enough about the particle physics in the next decade to permit accurate inferences about the rates of neutrino creation in the sun from the observed rates of neutrino arrival at the earth. The discussion in this subsection will presume that the required progress in understanding the properties of the neutrino is achieved.

●**Completing Hydrogen Fusion**

Table 1 shows that the two principal ways for completing nuclear fusion in the sun are reactions 3 and 4, the so-called $^3$He $-$ $^3$He and $^3$He $-$ $^4$He reactions. Because of the slightly smaller reduced mass that exists for the $^3$He $-$ $^3$He reaction, Coulomb barrier penetration favors this reaction over the $^3$He $-$ $^4$He reaction at lower temperatures. According to the standard solar model, the $^3$He $-$ $^4$He reaction is dominant in the innermost region of the sun (where it is 1.5 times faster than the $^3$He $-$ $^3$He reaction), but overall occurs in only about 19% of the fusion terminations that are described by Eq. (1). That is, in the most detailed solar models, the $^3$He $-$ $^3$He reaction is on the average more than five times faster in completing the nuclear fusion of protons into alpha particles than the competing $^3$He $-$ $^4$He reaction.

Is this prediction of the standard solar model correct? A determination of the $pp$ and $^7$Be neutrino fluxes (corrected for what non-standard particle physics has done to them after they were created in the sun) can answer this important question. The average ratio of the total number of $^3$He $-$ $^4$He reactions per unit time in the sun to the total number of $^3$He $-$ $^3$He reactions per unit time in the sun is

$$\frac{< \, ^3\mathrm{He} - {}^3\mathrm{He} \, >}{< \, ^3\mathrm{He} - {}^4\mathrm{He} \, >} \;=\; \frac{2\phi(^7\mathrm{Be})}{[\phi(pp) - \phi(^7\mathrm{Be})]}, \tag{8}$$

where $\phi(pp)$ and $\phi(^7\mathrm{Be})$ are, respectively, the fluxes from the $pp$ and $^7$Be neutrinos.

Eq. (8) is the most precisely-testable prediction that I know of from the theory of stellar energy generation. The known theoretical uncertainties in the calculation of the average solar ratio of $^3$He $-$ $^4$He to $^3$He $-$ $^3$He reactions is 7%.

### ●The $^8$B Neutrino Flux

The flux of neutrinos from $^8$B beta-decay in the sun (see reaction 9 of Table 1) is, in principle, the simplest solar neutrino flux to measure. The higher energies of the $^8$B neutrinos make them easiest to detect. For this reason, the Kamiokande, Superkamiokande, and SNO neutrino experiments will all concentrate on the $^8$B neutrinos.

However, one must determine the total flux of $^8$B neutrinos, including the more difficult to detect muon or tau neutrinos that may have been produced from the originally-created electron-type neutrinos. The total number of neutrinos of all types will be measured directly in the SNO experiment via the neutral-current disintegration of deuterium and, less directly, via electron-neutrino scattering in Superkamiokande. (This statement presumes there are no sterile neutrinos, see footnote 2 in §6.1.) In addition, one must determine precisely the low-energy cross section for the nuclear fusion reaction $p(^7\text{Be}, \gamma)^8\text{B}$, as discussed in §6.1.

The magnitude of the $^8$B flux (all neutrino flavors), which is a sensitive probe of the temperature of the solar interior, varies approximately as $\sim T_{(\text{central})}^{20}$. Therefore, it is important to determine experimentally the total $^8$B solar neutrino flux.

### ●The Temperature Profile of the Solar Interior

An accurate test of the theory of stellar structure and stellar evolution can be performed by measuring the average difference in energy between the neutrino line produced by $^7$Be electron capture in the solar interior and the corresponding neutrino line produced in a terrestrial laboratory. This energy shift is calculated to be 1.29 keV. The energy shift is approximately equal to the average temperature of the solar core, computed by integrating the temperature over the interior of the solar model with a weighting factor equal to the locally-produced $^7$Be neutrino emission. The total range of values for the shift, calculated for a number of modern solar models (going back to 1982), is 0.06 keV.

A measurement of the energy shift is equivalent to a measurement of the central temperature distribution of the sun.

The calculated energy profile of the $^7$Be line contains, analogous to line-broadening in classical (photon) astronomy, a description of the distribution of solar interior temperatures. The shape of the $^7$Be neutrino line is asymmetric: on the low-energy side, the line shape is Gaussian with a half-width at half-maximum of 0.6 keV and, on the high-energy side, the line shape is exponential with a half-width at half-maximum of 1.1 keV.

The calculated shape of the $^7$Be neutrino line is not affected significantly by vacuum neutrino oscillations, by the MSW effect, or by other frequently-discussed weak interaction solutions to the solar neutrino problems. This is a key result: it implies that the astronomical information contained in the line shift and in the line profile is not held hostage to further progress in neutrino physics.

Detectors are available that have the resolution to measure the line shift. Unfortunately, their current sizes are too small to permit a full-scale solar neutrino experiment. However, proposals have been made in the literature for developing detectors that are sufficiently large to be able to measure well the average shift in energy of the solar neutrino line.

### ●Ruling out 'Non-Standard' Solar Models

In the first decade and a half following the initial report that the measured solar neutrino flux in the chlorine experiment was less than the calculated value, a number of authors invented imaginative non-standard solar models that were designed to "solve" the solar neutrino problem. The situation now looks different. As described in §4, there are now three solar neutrino problems and it does not appear possible to reconcile the four operating experiments with any modification of stellar physics.

I think that the neutrino fluxes from the nuclear fusion reactions in the sun are known as well as the neutrino fluxes from many of the best terrestrial laboratory experiments. In the solar context, we use different constraints on the theoretical calculations than we do with laboratory accelerators. However, the solar constraints (especially the measured solar luminosity and the helioseismological frequencies) provide powerful limits on the allowed values of the neutrino fluxes. These constraints are discussed in §3.

Nevertheless, many physicists are unfamiliar with stellar physics. They do not feel comfortable judging the plausibility of different solar models, even fanciful ones in which the solar model contains, for example, a central black hole or a non-Maxwellian energy spectrum for the nuclei. Some physicists are willing to consider solar models that nearly all astrophysicists would dismiss as unworthy of discussion.

It would be instructive to calculate precise solar models based upon some of the more frequently discussed non-standard models (e.g., a low central heavy element abundance, iron precipitation, a very strong internal magnetic field, nearly complete element mixing, turbulent diffusion, massive mass loss, or energy transport by WIMPs). For each non-standard hypothesis, a solar model could be evolved using the best-available physics (opacities, equation of state, diffusion rates, and measured input data) while also imposing the *ad hoc* stellar structure hypothesis. The non-standard models computed in this way could be compared with the thousands of accurately-measured $p$-mode helioseismological frequencies.

I am confident that the non-standard models which have been suggested as possible solutions of the solar neutrino problem would be ruled out by accurate and detailed comparisons with helioseismological data. Most of the suggested models would fail, I suspect, on a grosser level, predicting for example the wrong depth of the convective zone or the wrong dependence of sound velocity on depth within the sun. But, it would be an important contribution to test the conjecture that the previously-suggested non-standard solar models that were concocted to solve the solar neutrino problem (when it was just one problem) all fail to account for well-established results of helioseismology.

### ●Discover the $g$-Modes

The most important discovery that one can anticipate being made in optical solar astronomy is the detection of the oscillations from gravity modes. Unlike the many pressure-mode ($p$-mode) oscillations that have been studied so far, the largest amplitudes of the $g$-modes occur in the solar interior; they are expected to be damped heavily in the outer regions of the sun. This concentration toward the center is particularly desirable if one wants to learn about solar interior properties that are relevant to neutrino astrophysics. However, the interior concentration also makes the detection of gravity modes very difficult. The amplitudes of the $g$-mode oscillations are expected to be very small on the surface of the sun, where they might be measured. There is no well-understood mechanism that predicts g-modes will carry enough energy to be observable. They have not yet been detected convincingly in the sun.

New experiments are underway to attempt to detect the $g$-modes from space (with the SOHO satellite) and from an international network of ground-based telescopes (GONG). The results of these new experiments will be of great interest for solar physics even if they do not lead to the detection of $g$-modes, since they will provide refined observations of the $p$-modes. If the $g$-modes are detected, it will be of epochal importance.

### ●More Complete Models of the Sun

The accuracy of the physical description that is currently achieved with one-dimensional (spherically symmetric) models of the sun that include diffusion is sufficient to permit excellent quantitative agreement with the measured $p$-mode oscillation frequencies. Numerical experiments and theoretical arguments also suggest that further improvements are unlikely to affect very significantly the calculated neutrino fluxes.

Nevertheless, current models of the sun are incomplete. They do not take account of the two-dimensional (or three-dimensional) nature of solar structure; they do not contain a self-consistent dynamical treatment of the effects of rotation, of magnetic fields, of mass loss, and of other possible effects that may violate the currently-used approximations of spherical symmetry and quasi-static evolution. We know observationally that the sun (at least near its surface) contains magnetic fields, that it is loosing mass, and that it departs from spherical symmetry by about 1 part in $10^5$.

There are both analytic and calculational challenges in including these complicated processes in a more complete physical description in the next generation of solar models. New self-consistent methods of calculating solar models (or stellar models) must be developed and then the appropriate numerical techniques must be worked out, tested, and applied.

The goal of developing a more complete solar model is a challenge for the next decade and beyond. Fortunately, it is a challenge that could lead to important progress since computing power is much greater than it was in the past and there is an abundance of precision data with which to make detailed comparisons.

Solar astrophysics has a bright future.

## 7. Summary

The first thirty years of solar neutrino research have verified experimentally the fundamental predictions of nuclear energy generation in stars. The next ten or twenty years of research will, I think, concentrate on using solar neutrinos to learn more about weak interaction physics. As the weak interaction questions are being resolved, it will be possible to carry out progressively more accurate tests of the theory of nuclear energy generation and of stellar structure.

In retrospect, the history of solar neutrino research seems ironic. It began with an effort to use neutrinos, whose properties were assumed well known, to study the interior of the nearest star. The project was an unconventional application of microscopic physics which was designed to carry out a unique investigation of a massive, macroscopic body, the sun. It now appears that a large community of physicists, chemists, astrophysicists, astronomers, and engineers working together may have stumbled across the first observed manifestation of physics beyond the standard electroweak model.

We may have been incredibly lucky.

Will the next generation of experiments show that physics beyond the electroweak model is definitely required to understand the solar neutrino experimental results? I do not know. But, I am sure that we have already learned important things about neutrino physics from the existing solar neutrino experiments and that we will learn additional things from the future experiments. This research may, or may not, lead to a consensus view that physics beyond the electroweak model is implied by solar neutrino experiments. To me the marvelous lesson of solar neutrino physics is that work on the forefront of one field of science can lead to important and completely unanticipated developments in a different field of science. This seems to me both humbling and beautiful.

## Bibliographic Notes

**1.** Recent descriptions of the ongoing solar neutrino experiments are contained in the following papers.

Davis, R. 1993, in Frontiers of Neutrino Astrophysics, eds. Y. Suzuki & K. Nakamura (Tokyo: Universal Academy Press), p. 47; Davis, R. 1994, Prog. Part. Nucl. Phys., 32, 13.

Suzuki, Y. 1995, Nucl. Phys. B (Proc. Suppl.) 38, 54; Proc. of the 6th Int. Workshop "Neutrino Telescopes", Venice, February 22-24, 1994 (ed. M. Baldo Ceolin), p. 197.

Abdurashitov, J. N., et al. 1994, Phys. Lett. B, 328, 234; Nico, G., et al. 1995, in Proceedings of the XXVII International Conference on High Energy Physics, July 1994, Glasgow, eds. P. J. Bussey and I. G. Knowles (Philadelphia: Institute of Physics), p. 965.

Anselmann, P., et al. 1994, Phys. Lett. B, 327, 377; *ibid*, 1995, 342, 440.

**2.** The standard model predictions used in this talk are taken from Bahcall, J. N., & Pinsonneault, M. 1995, Rev. Mod. Phys., October issue. The formulation of the neutrino problems is adapted from Bahcall, J. N. 1994, Phys. Lett. B, 238, 276.

**3.** For a general review of solar neutrino physics and astrophysics, the reader can consult two books devoted to the subject. *Neutrino Astrophysics* is a monograph by J. N. Bahcall that is published by Cambridge University Press (1989). *Solar Neutrinos: The First Thirty Years* contains reprints of 104 of the key papers in the development of the subject plus brief introductions and summaries of recent developments in the major subject areas: standard model expectations, solar neutrino experiments, nuclear fusion reactions, physics beyond the standard model, and helioseismology.

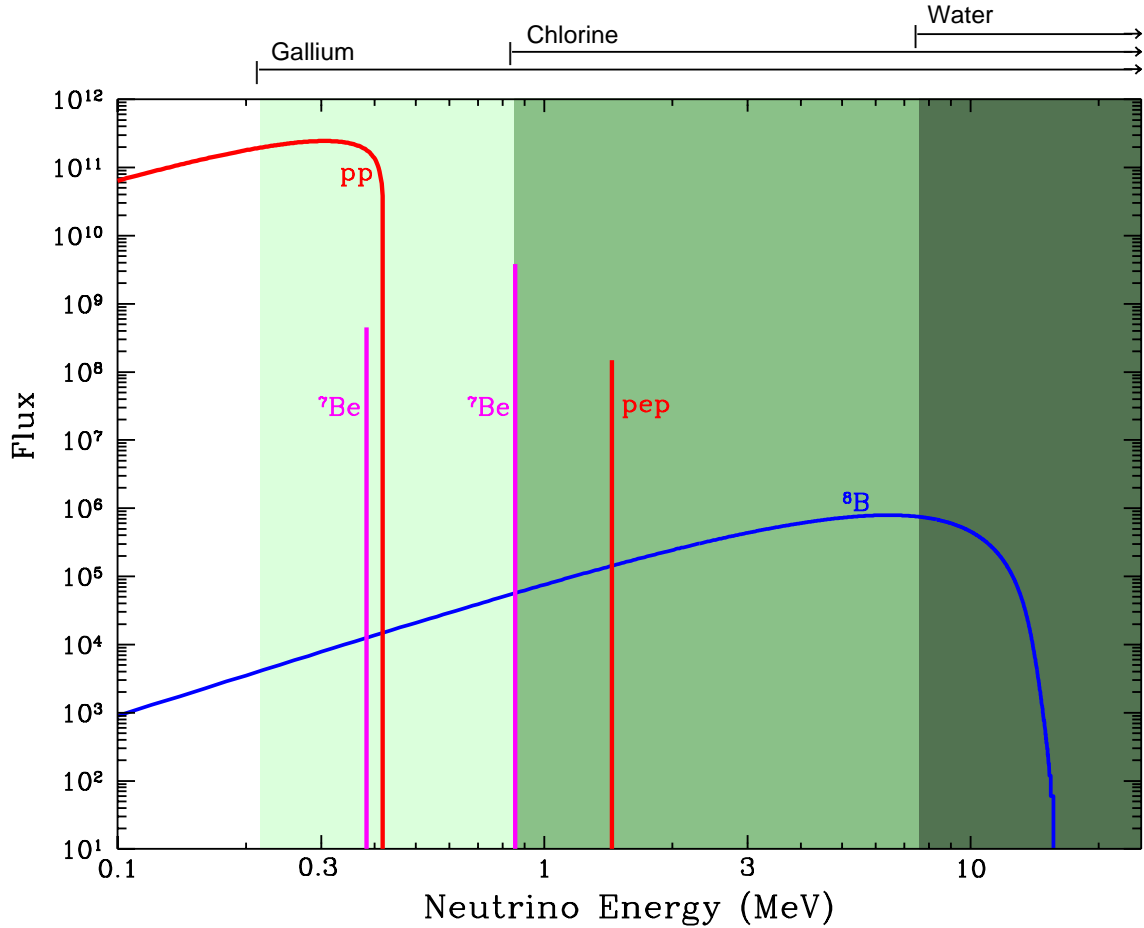

Fig. 1.— Solar Neutrino Spectrum. This figure shows the energy spectrum of neutrinos from the *pp* chain that is predicted by the standard solar model. The neutrino fluxes from continuum sources (*pp* and $^8$B) are given in the units of number per cm$^2$ per second per MeV at one astronomical unit. The line fluxes (pep and $^7$Be) are given in number per cm$^2$ per second. The arrows at the top of the figure indicate the energy thresholds for the ongoing neutrino experiments. The higher-energy $^7$Be line is just above threshold in the chlorine experiment.

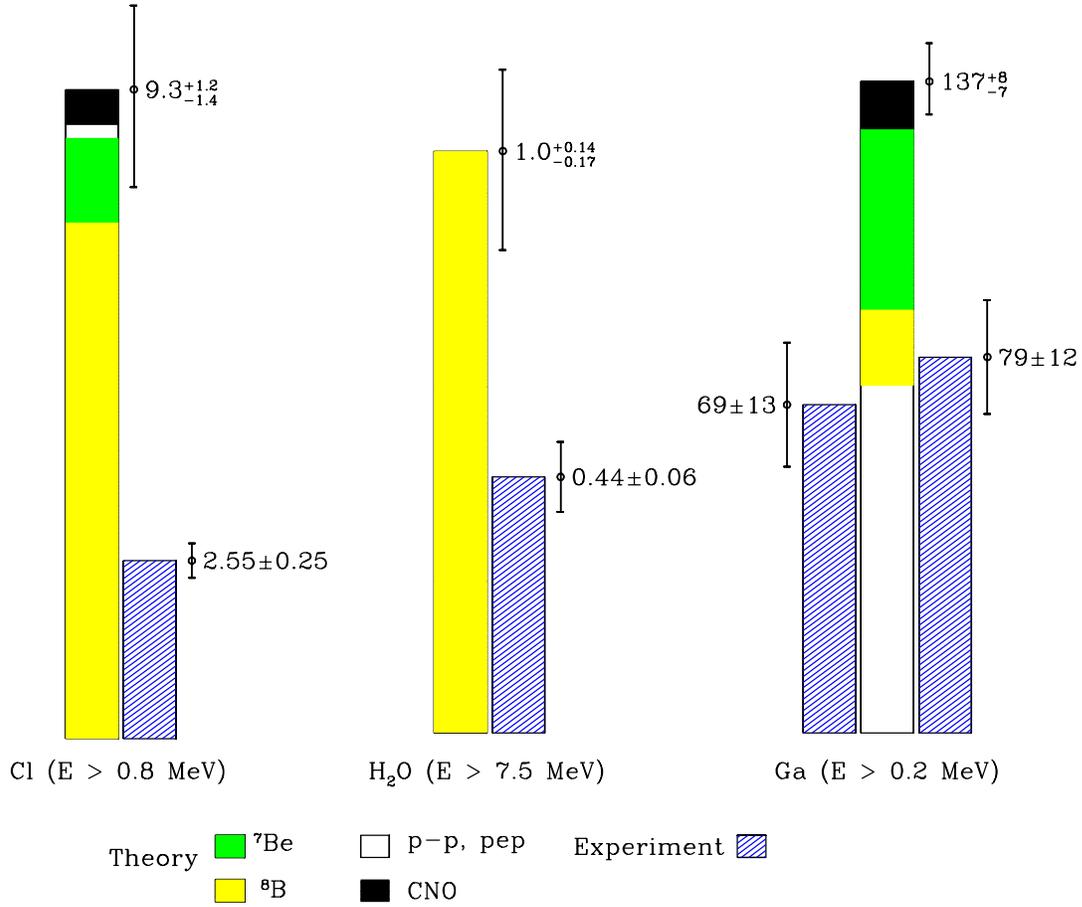

Fig. 2.— Comparison of measured rates and standard-model predictions for four solar neutrino experiments.